\documentclass[12pt]{article}

\usepackage{ragged2e} 
\usepackage{changepage}

\usepackage{siunitx} % degree symbol
\usepackage{amsmath}
\usepackage{amssymb}
\usepackage{xfrac} % for other fraction (e.g. sfrac)
\usepackage{bbm}

\usepackage{graphicx}
\usepackage{subcaption}
\usepackage{sidecap}
\usepackage{wrapfig}

\usepackage{caption}
\usepackage{verbatim}
\usepackage[english]{babel}

\usepackage{array}
\usepackage{adjustbox}
\usepackage{float}
\usepackage{enumerate}
\usepackage{multirow}
\usepackage{multicol}
\usepackage[table,xcdraw]{xcolor}
\usepackage[round]{natbib}

\usepackage{csvsimple}
\usepackage{bbm}
\usepackage{pdflscape}
\usepackage{breqn}
\usepackage{lscape}

\usepackage{amsfonts}
\usepackage{latexsym}
 \usepackage[titletoc]{appendix}  
\usepackage[colorlinks=true, citecolor= blue]{hyperref} % hyperlinks
\hypersetup{linkcolor=black} % hyperlinks
\usepackage[all]{hypcap} % hyperlinks

\usepackage[a4paper,top=2cm,bottom=2cm,left=3cm,right=3cm,marginparwidth=1.75cm]{geometry}

\usepackage{float} 
\usepackage{xcolor}
\usepackage{setspace}
\usepackage{hyperref}

\usepackage{authblk}

\newcommand{\shortsmile}[1]{\overset{\scalebox{0.6}[0.6]{$\smile$}}{#1}\mspace{-1mu}}

\title{\vspace{-1cm} \huge \textbf{Scaling Spatially Varying Distributed Lag Nonlinear Models for Nationwide Small-Area Data: A Divide-and-Conquer Approach}}
\date{}

\author[1]{Sara Rutten
\footnote{Corresponding author. {\textit{E-mail address}: sara.rutten@uhasselt.be}}}
\author[1,2]{Thomas Neyens}
\author[1]{Elisa Duarte}
\author[1]{Christel Faes}

\affil[1]{Interuniversity Institute for Biostatistics (I-BioStat), Data Science Institute (DSI), Hasselt University, Hasselt, Belgium}
\affil[2]{L-BioStat, Department of Public Health and Primary Care, KU Leuven, Leuven, Belgium}

\begin{document}
\maketitle

\begin{abstract}
Lagged and nonlinear exposure-response relationships are frequently modelled using distributed lag non-linear models (DLNMs). However, these relationships can vary substantially across locations due to differences in climate, population structure, adaptation and socio-economic characteristics, among others. Spatially-varying DLNMs can capture this heterogeneity, but their computational demands limit their application to fine-scale nationwide datasets with thousands of small areas.

In this paper, we adapt a `divide-and-conquer´ approach to spatially varying DLNMs. The method partitions the spatial domain into smaller regions, expands each region with a $k$-order neighbourhood to reduce boundary effects, fits separate models within each partition using Laplacian P-splines, and combines the resulting estimates using conditional predictive ordinate–based mixture weights. 

We evaluated the approach in a simulation study based on municipalities in Sicily and applied it to daily summer temperature and mortality data from 7,122 municipalities in mainland Italy between 2011 and 2021. Simulation studies show good performance, with relative-risk estimates comparable to those from a global model and relatively robust to the choice and number of partitioning units. Computational efficiency can be substantially improved by using parallel computations.
The Italian application reveals significant spatial heterogeneity in the temperature-mortality relationship, with clear regional differences.  

The proposed framework makes spatially varying DLNMs feasible for large, fine-scale spatiotemporal datasets while retaining location-specific exposure– lag–response estimation.

\noindent \textbf{Keywords: Spatial heterogeneity, Distributed lag non-linear models, Laplace approximations, big-data analysis, small-area analysis} 
  
 \thispagestyle{empty}
\end{abstract}

\newpage
\clearpage\null

\pagenumbering{arabic}

\section{Introduction}
Assessing the impact of environmental conditions on population-level outcomes is a central challenge in many scientific disciplines. Quantifying the effect of these exposures often requires accounting for both immediate and long-term influences. A correct understanding of these delayed relationships is crucial, as it enables a more accurate estimation of the cumulative burden associated with the exposures. To flexibly model these delayed exposure effects, \cite{gasparrini2010} introduced the distributed lag non-linear models (DLNMs), providing a powerful framework to account for non-linear exposure and lagged effects, through the use of splines. This model has become widely used during the last years. For example, DLNMs have been used to investigate the delayed associations between temperature and mortality \citep{Gasparrini2015, Bao2016}, air pollution and COVID-19 incidence \citep{Rutten2025}, temperature and thermal tolerance in mayflies \citep{Redana2024} and drought and vegetation health \citep{Xu2024}. 

Several extensions of this model have been proposed throughout the years. Penalised formulations have been proposed to further increase the flexibility of DLNMs while controlling overfitting \citep{Gasparrini2017,Eilers1996}. \cite{Rutten2026} embedded this penalized DLNM  within a Bayesian estimation framework for use with small-area count data. However, these models assume spatial homogeneity of the exposure effect, an assumption often unrealistic with a large number and heterogeneous regions. Spatially varying DLNMs address this limitation by allowing the exposure–lag–response relationship to differ across locations \citep{Rutten2026_spatial}. Compared with conventional two-stage meta-analysis approaches \citep{Gasparrini2013}, this model is particularly suitable for small-area analyses in which outcome counts may be sparse and spatial dependence is important. 

The flexibility of spatially varying DLNMs comes at a substantial computational cost. Although the model can be fitted to datasets containing several hundred areas, memory requirements and computation time increase rapidly with the number of locations. As a result, fitting spatially-varying DLNMs may be infeasible for nationwide datasets with thousands of areas. This limits the use of spatially varying DLNMs in high-resolution studies, despite the growing availability of detailed spatiotemporal health and environmental data. In particular, jointly accounting for complex lag structures, spatial heterogeneity and large scale data remains computationally challenging.

Several approaches have been proposed to improve the scalability of spatio-temporal models. To scale spatio-temporal disease mapping to big-data problems, \cite{Lee2025} introduced an observation-driven process, augmenting the autoregressive type random effects in the model with functions of earlier data as additional covariates. An alternative is the `divide-and-conquer´ method proposed by \cite{Orozco-Acosta2023}. They divide the spatial domain into different spatial subdomains, fitting separate models in each partition. To avoid border effects, neighbouring areas (forming a `border´) are added to the partitions. Afterwards, the results are merged to obtain marginal posterior distributions of the relative risks (RRs), based on for example a mixture distribution approach. Although this method has not been formulated in the context of DLNMs, it provides a promising framework for extending these models to much larger datasets. For example, such an extension is especially relevant for temperature-mortality studies, where DLNMs are highly adequate for capturing the delayed and non-linear effects of temperature. A scalable spatially varying DLNM method would allow these complex associations to be modelled, while retaining spatial variation across thousands of areas. This is particularly important for nationwide studies, enabling the researchers to uncover localized vulnerabilities en heterogeneities that cannot be discovered in aggregated analyses, resulting in more detailed insights for public health interventions. 

%The increasing availability of high-resolution spatio-temporal data offers opportunities to thoroughly study environmental risks at a fine geographical scale, but also poses substantial computational challenges. Methods that can efficiently use these data, while preserving model flexibility are therefore essential. In particular, 

In this study, we adapt the `divide-and-conquer' framework of \cite{Orozco-Acosta2023}  to the spatially varying DLNMs proposed by \cite{Rutten2026_spatial}. This extension enables municipality-level estimation of heterogeneous temperature–mortality associations in nationwide datasets that cannot be fitted using a single global model.  This combination allows us to retain the flexibility of spatially-varying DLNMs, while substantially improving computational scalability, enabling a big-data spatially-varying temperature-mortality study of Italy. Section~2 describes the dataset and outlines the proposed methodology. Section 3 evaluates the accuracy and computational performance of the approach in a simulation study based on municipalities in Sicily. The method is applied to daily summer temperature and mortality data from 7,122 municipalities in mainland Italy between 2011 and 2021 in Section 4.

\section{Material and methods}
\subsection{Data}
Daily data on municipality-level mortality in $7,122$ municipalities in mainland Italy were retrieved from the Italian National Institute of Statistics (ISTAT) for the period 2011-2021 \citep{Istat_DecessiCauseMorte_2022}. The city boundaries and municipality-specific population size, as defined in 2021, were downloaded from ISTAT  \citep{ISTAT_AdminBoundaries2018} and were used to define the spatial units and population offsets in the statistical model. Daily temperature data during the summer period, defined from May to September, from 2011 to 2021 were extracted from the fifth generation of the European Reanalysis (ERA5) dataset \citep{ERA5LandDailyStats2025}. This dataset provided daily maximum temperature data at 2 m above the land surface, on a grid of approximately $0.25^\circ$. These data were aggregated to the municipality level by averaging the temperature values of all grid cells for which the centroid falls within the municipality boundary. 

Italy is characterized by large climatic differences, with a generally cooler climate in the northern areas and a Mediterranean climate in the south \citep{Fratianni2017}, and topographic differences further influence the temperature distribution. Moreover, Italy shows demographic heterogeneity in population structure, ageing and socio-economic composition \citep{Armenise2024}.

\subsection{Statistical methods}
\subsubsection{Spatially varying DLNM}
Differences in climate, population characteristics and possible adaptation to heat \citep{Masselot2025} make the assumption of spatial homogeneity of the temperature effect across Italy implausible. Therefore a spatially-varying DLNM is required to allow the temperature–mortality association to differ among municipalities. 

Denote by $y_{t,j}$ and $x_{t,j}$ the mortality count and maximum temperature at time $t = 1,\ldots, 1683$ and municipality $j=1,\ldots,7122$ respectively. Defining $\mu_{t,j}=\mathbb{E}(y_{t,j})$ and assuming the counts follow a Poisson distribution, the model was specified as:
\begin{align*}
    \log(\mu_{t,j}) = &\beta_0+\text{ns}(\text{year}_t,\text{df}=3; \boldsymbol{\gamma_r})+\text{ns}(\text{month}_t,\text{df}=2; \boldsymbol{\kappa_r})+\sum_{k=1}^{6}{\delta_{k} w_{k,t}} \\ + &s(x_{t,j}\ldots x_{t-L,j};\boldsymbol{\theta})+s(x_{t,j}\ldots x_{t-L,j};\boldsymbol{\theta_j})+u_j + \log(N_{j}).
\end{align*}
where $\beta_0$ is the intercept, $w_k$ is an indicator of the day of the week, $N_j$ the population size of municipality $j$. Year and month were modelled as a natural spline with $3$ and $2$  degrees of freedom respectively, to account for long-term and seasonal temporal variation. The municipality-level random effect $u_j$ accounts for residual spatial variation and was assigned a Leroux conditional autoregressive structure \citep{leroux1999}, $$\boldsymbol{u}\sim \mathcal{N}(\boldsymbol{0},\sigma^2[\rho Q + (1-\rho)I]^{-1}),$$ where $Q$ represents the neighbourhood matrix, $I$ the identity matrix, $\sigma^2$ the variance and $\rho \in [0,1]$ the parameter controlling the strength of spatial autocorrelation. The temperature association was modelled over a maximum lag of $L=8$ days. The term $s(x_{t,j}\ldots x_{t-L,j};\boldsymbol{\theta})$ represents the exposure–lag–response surface shared across all municipalities, whereas $s(x_{t,j}\ldots x_{t-L,j};\boldsymbol{\theta_j})$ represents the municipality-specific deviation from the common surface. 

Following \cite{Rutten2026_spatial}, we can write out the two smooth functions using tensor-product basis functions for the exposure and lag dimensions, such that:
\begin{align*}
    \log(\mu_{t,j}) = &\beta_0+\text{ns}(\text{year}_t,\text{df}=3; \boldsymbol{\gamma_r})+\text{ns}(\text{month}_t,\text{df}=2; \boldsymbol{\kappa_r})+\sum_{k=1}^{6}{\delta_{k} w_{k,t}} \\ &+ \sum_{i=1}^{v_x} \sum_{k=1}^{v_l}\left(\sum_{l=0}^L\widetilde{b}_i(x_{t-l,j})\shortsmile{b}_k(l) \right)(\theta_{ik}+\theta_{j,ik})+u_j + \log(N_{j}).
\end{align*}
where $\{\widetilde{b}_i\}_{i=1}^{v_x}$ and $\{\shortsmile{b}_k\}_{k=1}^{v_l}$ are basis functions for the exposure and lag dimension, respectively. In this formula, $\theta_{ik}$ denotes the coefficient of the common exposure–lag–response surface, whereas $\theta_{j,ik}$ represents the deviation of municipality $j$ from that common surface.  Consequently, the municipality-specific surface is determined by the sum $\theta_{ik}+\theta_{j,ik}$. We used the Type IV prior specification of \cite{Rutten2026_spatial}, which combines penalized P-splines with spatially structured municipality-specific deviations. Consequently, neighbouring municipalities are encouraged to have similar deviations, while the estimated relationships remain smooth across the exposure and lag dimensions. Separate smoothing parameters control the common surface and the municipality-specific deviations. Model parameters were estimated using Laplacian P-spline estimation. Conditional on the hyperparameters, the posterior distribution was approximated by a multivariate Gaussian distribution centred at the posterior mode, and the smoothing, precision and spatial-correlation hyperparameters were estimated through numerical optimisation. Full details of the prior specification and estimation are provided in \cite{Rutten2026_spatial} and \cite{gressani2018}.

The impact of an exposure $x$ as compared to a reference exposure $x_0$ in municipality $j$ is then quantified as the lag-specific log relative risk:
\begin{align*}
   \log\bigl(RR_{x,x_0,j}(l)\bigl) &=\sum_{i=1}^{v_x}\sum_{k=1}^{v_l}{\bigl(\widetilde{b}_i(x)-\widetilde{b}_i(x_0)\bigl)\shortsmile{b}_k(l)(\hat{\theta}_{ik}+\hat{\theta}_{j,ik})}, 
\end{align*}
or the overall cumulative log relative risk over lags $0,\ldots, L$:
\begin{align*}
     \log\bigl(RR_{x,x_0,j}^{\text{overall}}\bigl) &=\sum_{i=1}^{v_x}\sum_{k=1}^{v_l}\biggl(\sum_{l=0}^{L}{\bigl(\widetilde{b}_i(x)-\widetilde{b}_i(x_0)\bigl)\shortsmile{b}_k(l)\biggl)(\hat{\theta}_{ik}+\hat{\theta}_{j,ik})}. 
\end{align*}

\subsubsection{Divide-and-conquer method}
With long time series from $7122$ municipalities, this model is computationally infeasible because of the dimension of the municipality-specific spline effects and the associated spatial precision matrices. Therefore, we adapt the `divide-and-conquer´ method of \cite{Orozco-Acosta2023} to the spatially varying DLNM framework. 

We first divide the complete spatial domain into a set of non-overlapping partitions. To diminish boundary effects, each partition is enlarged by adding all municipalities within a $k$-order neighbourhood of the municipalities in that partition. As a result, municipalities near a partition boundary could be included in more than one enlarged partition.  A separate Type IV spatially varying DLNM, as described in Section 2.2.1, was then fitted to each enlarged partition using Laplacian P-spline estimation. 

For municipalities appearing in only one submodel, inference was based directly on the posterior distribution obtained from that submodel. For municipalities included in multiple partitions,  the posterior distributions from different submodels are combined using a mixture strategy \citep{Orozco-Acosta2023}. More specifically, the posterior distribution of the relative risks (RRs) is formed by combining the individual posterior density functions with weights proportional to the conditional predictive ordinate (CPO) \citep{Congdon2020}. In contrast to the model of \cite{Orozco-Acosta2023}, our RR estimates are only spatially-varying but remain time-invariant within this spatio-temporal model. Therefore, to obtain area-specific weights, we aggregate the time-indexed CPO values by taking their product over the entire time window. The posterior density $f(x)$ of the RR in area $j$ at a specific temperature, is then formed by: 
\begin{align*}
    w_{s} &= \frac{\text{CPO}_{sj}}{\sum_{s=1}^{m(j)}{\text{CPO}_{sj}}} \\
    f(x) &= \sum_{s=1}^{m(j)}{w_{s}f_s(x)} 
\end{align*}
with $m(j)$ the number of submodels in which area $j$ occurs, $f_s(x)$ the density function in submodel $s$ and $\text{CPO}_{sj}$ the corresponding (aggregated) CPO value of area $j$.
\section{Simulation}
In this simulation study, we aim to investigate the impact of the choice of partition-units, including both the number of partitions as well as the partitioning design, e.g. random partition or administrative boundaries. Since we also want to compare the accuracy we obtain using this `divide-and-conquer´ method with the accuracy obtained using one global model, we cannot perform this simulation study on municipality-level Italy due to memory and time constraints. Therefore, we performed the simulation study on Sicily, since \cite{Rutten2026_spatial} showed that a global model can be estimated on this data. 

\subsection{Simulated data}
The time-varying predictor $x_{t,j}$ is the observed daily summer temperature  in all municipalities of mainland Sicily  ($j=1,\ldots,383$) between 2011 and 2021, extracted from the fifth generation of the European Reanalyses (ERA5) dataset \citep{ERA5LandDailyStats2025}. These data were standardized to the interval from $0$ to $10$. 

An area-specific daily count $y_{t,j}$ was simulated from a Poisson distribution with mean:
\begin{align*}
    \log(\mu_{t,j}) = \beta_0 + \sum_{l=0}^{8}{f_j \cdot w_j(x_{t-l,j},l)+u_j+\log(N_j})
\end{align*}
with $u_j$ a Leroux random effect with $\rho = 0.90$ and $\sigma^2 = 0.05$. The intercept $\beta_0=-10$ such that the outcome represents a realistic mortality rate. The population size $N_j$ is the true municipality-specific population size of Sicily in $2021$. The function  $f_j \cdot w_j(x_{t-l,j},l)$ describes the area-specific RR-curve. We assumed an underlying east-to-west gradient, with in general, higher cumulative RRs in the west compared to the east (Figure~\ref{fig:simulation_setting}). More information can be found in the Supplementary Materials.
\begin{figure}
    \centering
    \includegraphics[width=0.8\linewidth]{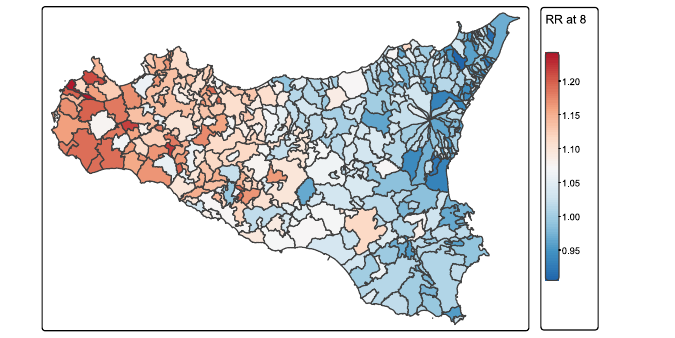}
    \caption{\emph{Simulated overall cumulative RR (over lag $0-8$) at exposure level 8, according to an east to west gradient.}}
    \label{fig:simulation_setting}
\end{figure}
\subsection{Fitted models and partitioning strategies}
To investigate the impact of the partitioning structure, we compared models using $5$ different partition strategies. The partitioned models differed in the geometry of their core partitions: (1) Global model (no partitioning), (2) Province-based partitioning ($9$ provinces), (3) Regular grid partitioning (both row and column wise), (4) North-to-south row partitioning and (5) East-to-west column partitioning. Partitioning structure (3) to (5) were constructed using the random partitioning method of \cite{Orozco-Acosta2023} with a minimum and maximum required size of $25$ and $200$ municipalities per region respectively. A regular grid ($7 \times 7$), row grid ($7 \times 1$) and column grid ($1 \times 7$) were used to construct these different partitions. 

\begin{figure}
\centering

\begin{minipage}{0.45\textwidth}
\includegraphics[width=\linewidth]{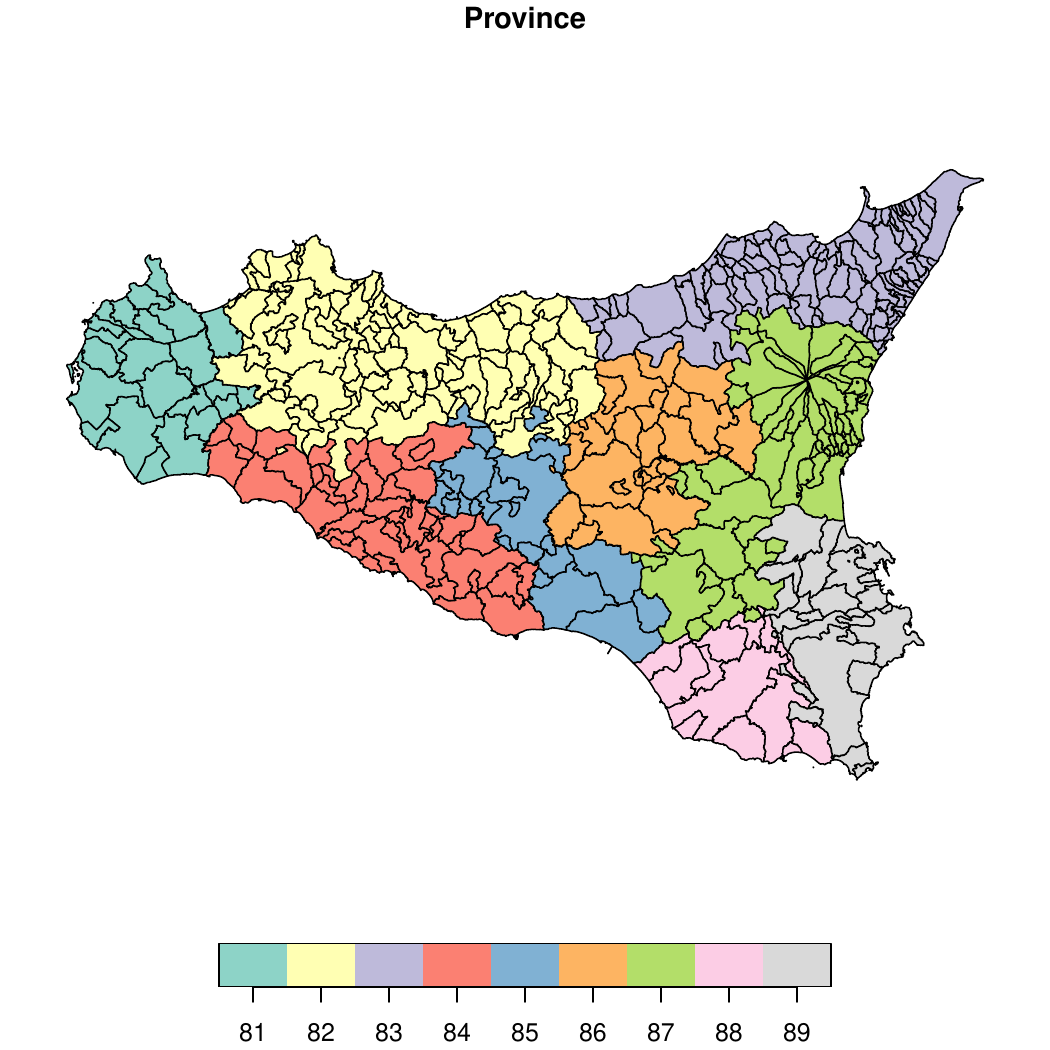}
\end{minipage}
\hfill
\begin{minipage}{0.45\textwidth}
\includegraphics[width=\linewidth]{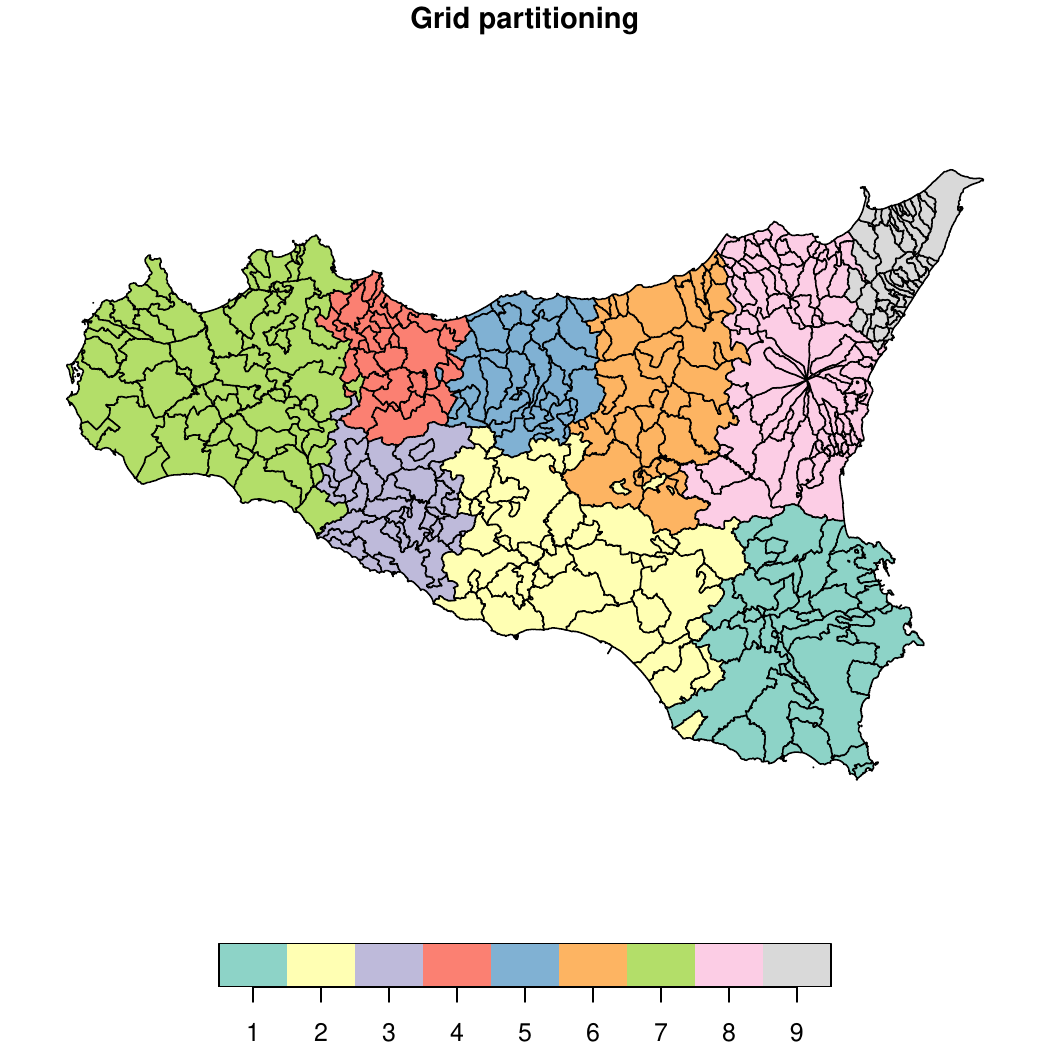}
\end{minipage}

\vspace{1cm}

\begin{minipage}{0.45\textwidth}
\includegraphics[width=\linewidth]{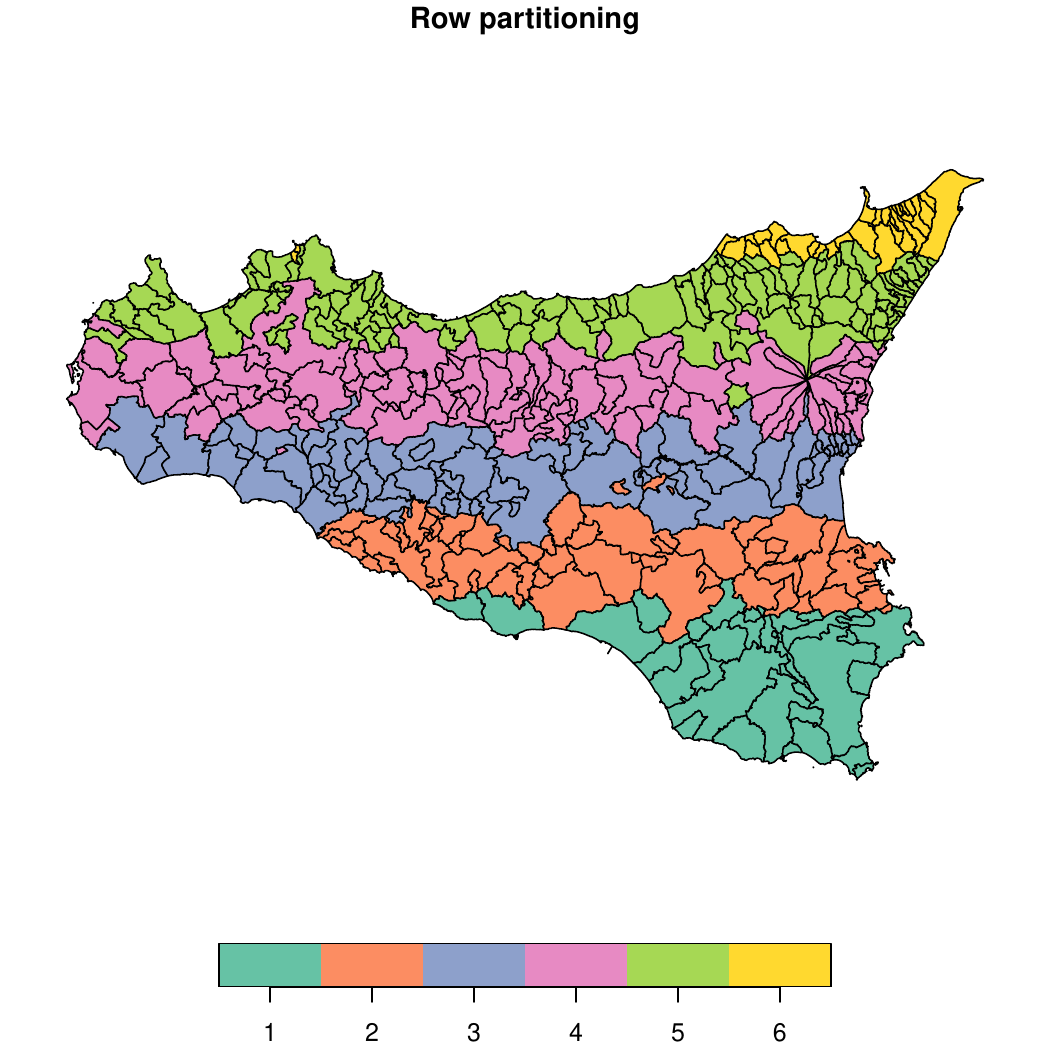}
\end{minipage}
\hfill
\begin{minipage}{0.45\textwidth}
\includegraphics[width=\linewidth]{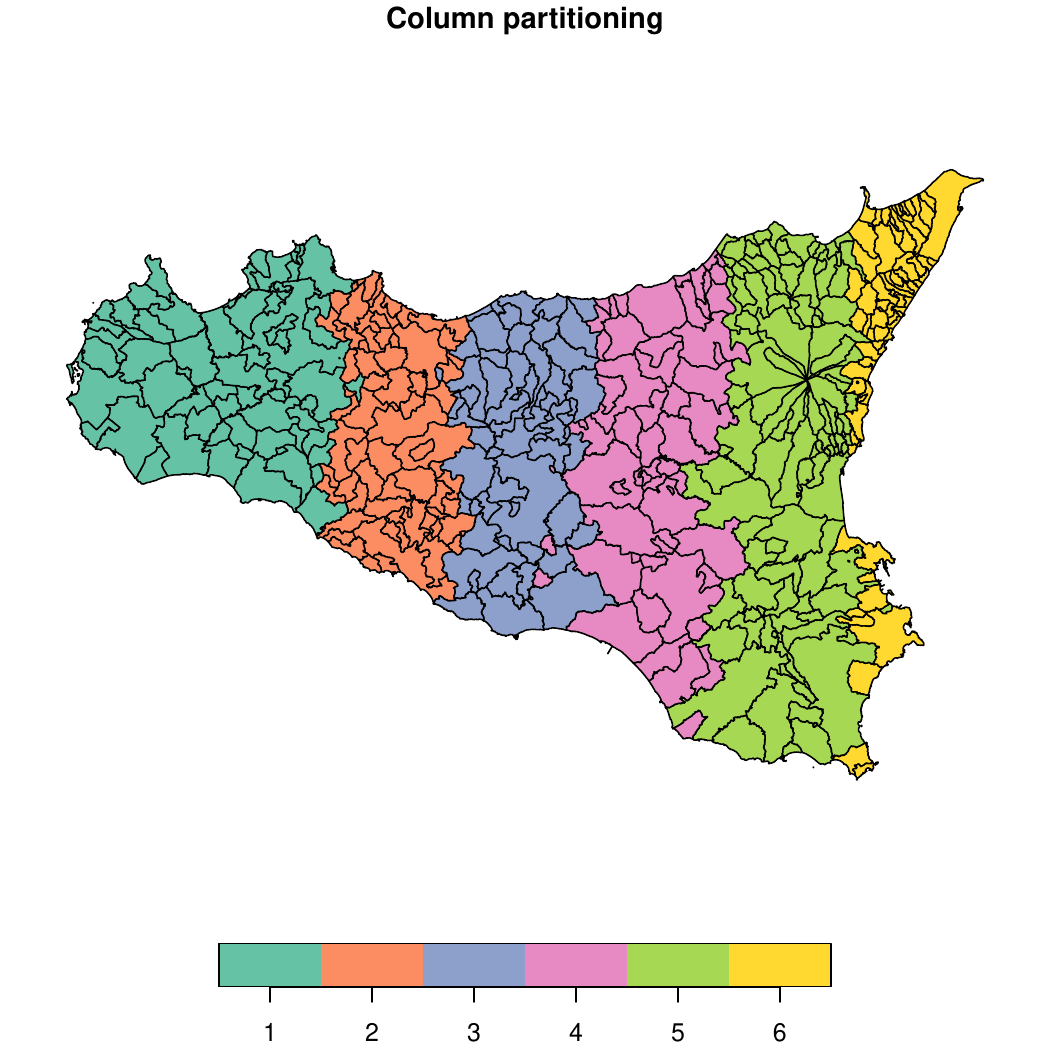}
\end{minipage}

\vspace{1cm}

\begin{minipage}{0.45\textwidth}
\includegraphics[width=\linewidth]{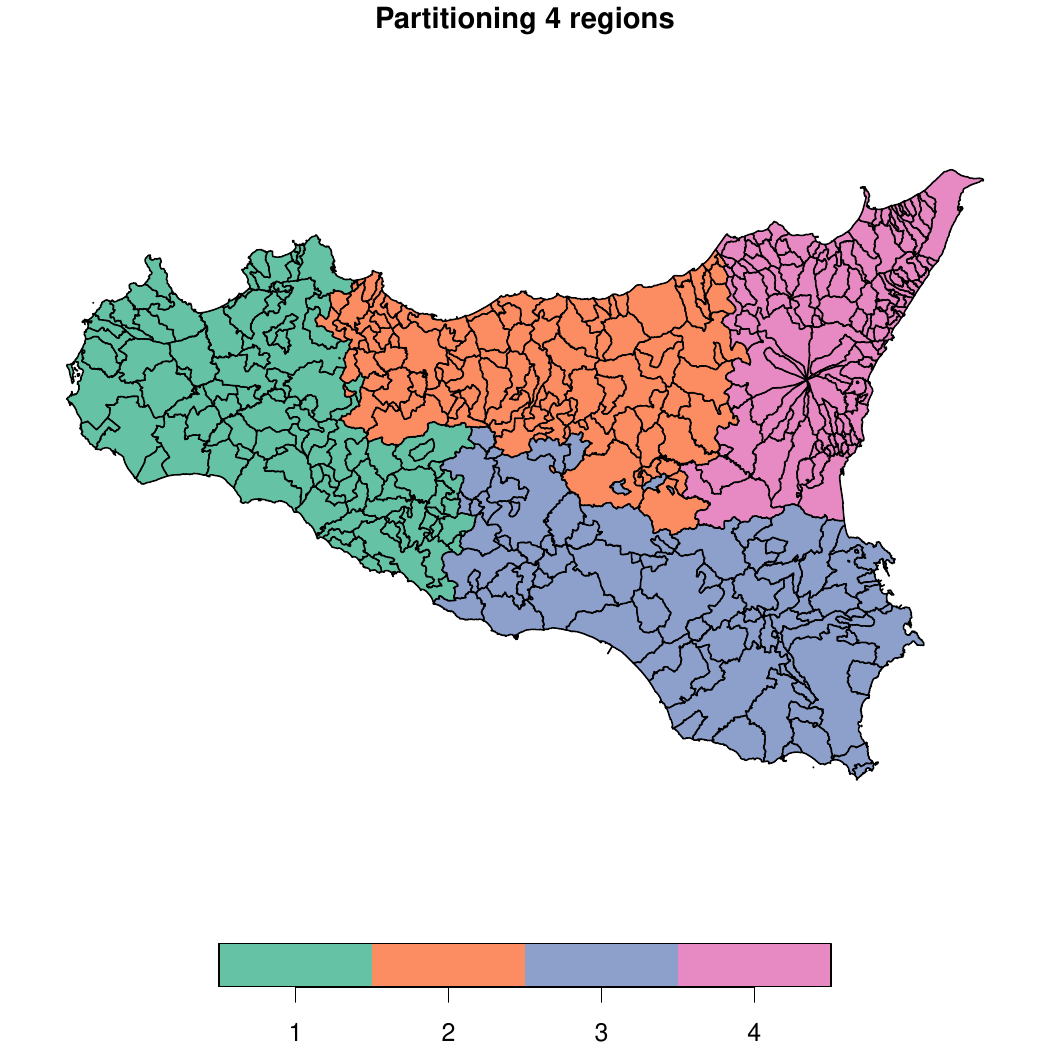}
\end{minipage}
\hfill
\begin{minipage}{0.45\textwidth}
\includegraphics[width=\linewidth]{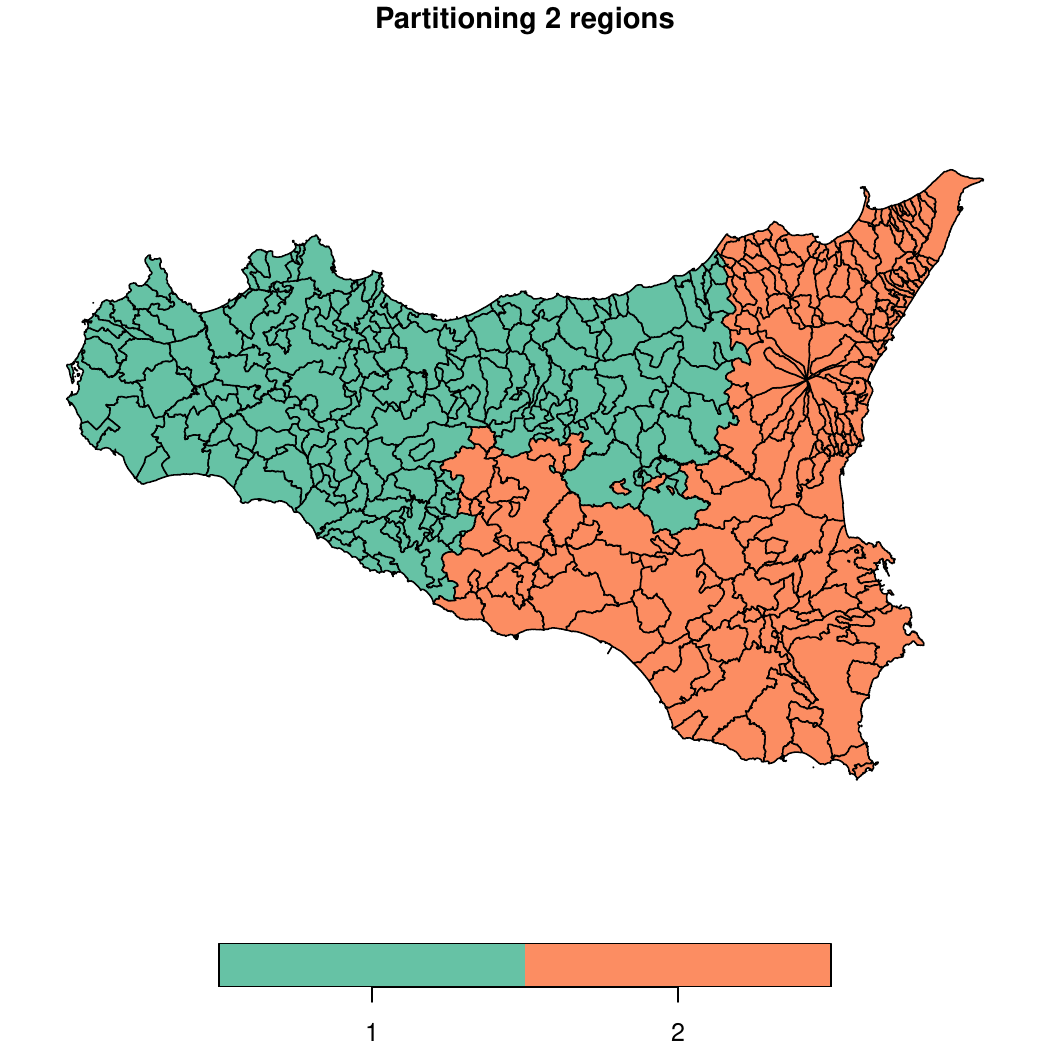}
\end{minipage}
\caption{\emph{Partitioning structure of the different models (2) to (7) of the simulation study.}}
\label{fig:structure_partitioning}
\end{figure}

To investigate the effect of the number of partitions on the computation time, we additionally compared regular-grid structures containing different numbers of partitions: (1) Complete model and (3) Regular grid partitioning with $9$ regions with (6) Regular grid partitioning with $4$ regions and (7) Regular grid partitioning with $2$ regions. These last two partitioning structures are defined by combining multiple regions from structure (3). The $6$ different partitioning structures are visualized in Figure~\ref{fig:structure_partitioning}. 

Each core partition was enlarged by including municipalities within its second-order neighbourhood to reduce boundary effects. A separate Type IV spatially varying DLNM was fitted to every enlarged partition using Laplace approximations. For each model, we used penalized splines with $6$ degrees of freedom in both dimensions and a maximum lag of $8$. A Leroux structure was assumed on the area-specific random effect as well as on the area-specific deviations from the overall exposure-lag-response curve.

\subsection{Model performance}
For municipality $j$, we evaluated the lag-specific log relative risk at exposure $x$, relative to reference exposure $x_0$ 
and the overall cumulative log relative risk over lags $0,\ldots, L$.
We used $x_0 = 5$ as the reference exposure and $x$ evaluated relative risks on a grid of exposures $x=0,1,\ldots,10$. Estimation accuracy was assessed using the RMSE and $95\%$ coverage for both the lag-specific and overall cumulative relative risks (more information can be found in the Supplementary Materials). 

We ran $100$ simulations for each partitioning structure, using R on the Flemish Supercomputer (VSC). We reported both the computation time required to fit all submodels, including the calculation of the time-indexed CPO-values, and the prediction time, including the construction of the submodel-specific mixture weights, in minutes. 

\subsection{Simulation Results}
Table \ref{tab:simulation_study1} summarizes the effect of different partitioning structures on the accuracy of the model. Partition geometry had little effect on estimation accuracy. Both in terms of coverage and RMSE, partitioning methods (2) to (5) perform similarly compared to the global model. There was no consistent advantage for province-, grid-, row-, or column-based partitioning. In particular, although an underlying east to west gradient is assumed, the model using an east to west partition (method (5)) does not show a superior performance compared to the other methods. Empirical coverage was slightly higher for the partitioned models than for the global model, but their credible intervals were also wider. This indicates that the divide-and-conquer approach preserved point-estimation accuracy while producing somewhat more conservative uncertainty estimates. Computation times are reported using a sequential fitting procedure for the `divide-and-conquer´ method. In a setting that global model fitting is computationally feasible, fitting one global model has significantly lower computation time compared to the partitioning methods. 

Not only the partitioning structure but also the number of partitions influences the required computation time. Table \ref{tab:simulation_study2} compares the computation time of 3 grid structures, differing in the number of partitioning units. We reported the computation time using both a sequential fitting procedure and parallel method (allowing for $2$ or $3$ models to be fitted in parallel). The total estimation time, prediction time and minimum, maximum and median estimation time of the submodels can be found in the table. A higher number of partitioning units results in smaller computation times of the partition-specific submodels but, as more models are required, the overall estimation time does not necessarily decrease, especially when using a sequential fitting procedure. When allowing for parallel computations, the overall computation time does decrease drastically. However, as illustrated by the $2$-regions grid (7), parallel computations can increase the computation time of the individual submodels, likely due to memory constraints. 

Because the primary simulation scenario considered a smooth east-to-west gradient, we also conducted a second simulation study with a more discontinuous coast-to-inland trend to assess the robustness of the results. As similar conclusions could be drawn, the results are reported in the Supplementary Materials. 

Overall, the different partitioning structures do not seem to highly influence the obtained accuracy of the RRs. However, the `divide-and-conquer´ method results in slightly wider credible intervals, compared to global models. It can be seen that partitioning does not necessarily decrease computation time, especially when using sequential fitting. However, in big-data studies, fitting a global model is often impossible due to memory constraints. Choosing the number and structure of partitions, is in real-life limited by practical, time and memory constraints. 

\begin{table}[h]
\caption{\emph{Simulation results for comparing partitioning structure (1) to (5): RMSE of lag-specific RR (RMSE RR), RMSE of cumulative RR, over lag $0-8$, (RMSE RR overall), coverage of lag-specific RR (cov RR),  coverage of cumulative RR (cov RR overall). The computation time and prediction time are reported in minutes.}}
\label{tab:simulation_study1}
\centering
\begin{adjustbox}{width=\linewidth} % Optional for scaling if necessary
\begin{tabular}{ll|ccccc}
\hline
 &  & \textbf{(1) Global} & \textbf{(2) Prov} & \textbf{(3) Grid 9 regions} &\textbf{(4) N-S} & \textbf{(5) E-W}  \\ 
 \hline
 \multicolumn{2}{l|}{Cov RR} & $0.9967$& $0.9988$& $0.9989$ & $0.9981$& $0.9988$  \\
  \multicolumn{2}{l|}{Length CI RR} & $0.0553$& $0.0659$& $0.0663$ & $0.0620$& $0.0625$  \\
\multicolumn{2}{l|}{Cov RR overall} & $0.9822$ & $0.9915$& $0.9919$ & $0.9854$ & $0.9918$   \\
\multicolumn{2}{l|}{Length CI RR} & $0.2904$ & $0.3168$& $0.3170$ & $0.3088$ & $0.3053$   \\
\multicolumn{2}{l|}{RMSE RR} & $0.0121$& $0.0124$& $0.0126$ & $0.0125$ & $0.0120$ \\
\multicolumn{2}{l|}{RMSE RR overall} & $0.0734$ & $0.0699$ & $0.0705$ & $0.0757$ & $0.0671$ \\
\multicolumn{2}{l|}{Computation time} & $10.83$ & $23.75$ & $22.31$ & $27.30$ & $20.58$  \\
\multicolumn{2}{l|}{Prediction time} & $2.98$ & $2.31$ & $2.03$ & $4.21$ & $2.40$  \\
\hline
\end{tabular}
\end{adjustbox}
\end{table}

\begin{table}[h]
\caption{\emph{Simulation results for comparing computation time partitioning structure (1), (3), (6) and (7): RMSE of lag-specific RR (RMSE RR), RMSE of cumulative RR, over lag $0-8$, (RMSE RR overall), coverage of lag-specific RR (cov RR),  coverage of cumulative RR (cov RR overall). The computation time is reported in minutes. We reported the total computation time, the prediction time and the minimum, maximum and median estimation time of the submodels. }}
\label{tab:simulation_study2}
\centering
\begin{adjustbox}{width=\linewidth} % Optional for scaling if necessary
\begin{tabular}{lll|cccc}
\hline
 & & & \textbf{(1) Full} & \textbf{(3) Grid 9} &\textbf{(6) Grid 4} & \textbf{(7) Grid 2}  \\ 
  & & &  & \textbf{regions} &\textbf{regions} & \textbf{regions}  \\ 
 \hline
 \multicolumn{3}{l|}{Cov RR} & $0.9967$& $0.9989$& $0.9986$ & $0.9979$ \\
\multicolumn{3}{l|}{Cov RR overall} & $0.9822$ & $0.9919$& $0.9913$ & $0.9882$    \\
\multicolumn{3}{l|}{RMSE RR} & $0.0121$& $0.0126$& $0.0120$ & $0.0118$ \\
\multicolumn{3}{l|}{RMSE RR overall} & $0.0734$ & $0.0705$ & $0.0683$ & $0.0697$ \\
\hline
\textbf{Fitting} & \textbf{Computation time} & &&&&\\
\hline
\multirow{4}{*}{Sequential} & \multicolumn{2}{l|}{Total time} & $11.16$ & $23.25$ & $18.43$ & $15.89$  \\
& Partition time & Min & $\cdot$ & $1.80$& $3.76$ & $7.43$ \\
& & Median & $\cdot$& $2.26$& $4.40$ & $\cdot$  \\
& & Max & $\cdot$& $3.99$& $5.35$ & $8.03$  \\
\hline
\multirow{4}{*}{2 parallel} & \multicolumn{2}{l|}{Total time} & $\cdot$ & $14.33$ & $12.38$ & $14.70$  \\
& Partition time & Min & $\cdot$  & $1.86$& $3.99$ & $10.95$ \\
& & Median & $\cdot$ & $2.28$& $4.52$ & $\cdot$  \\
& & Max & $\cdot$ & $4.08$& $5.72$ & $12.91$  \\
\hline
\multirow{4}{*}{3 parallel} & \multicolumn{2}{l|}{Total time} & $\cdot$ & $10.82$ & $10.77$ & $\cdot$   \\
& Partition time & Min & $\cdot$  & $1.86$& $4.13$ & $\cdot$  \\
& & Median &$\cdot$  & $2.38$& $4.59$ & $\cdot$   \\
& & Max & $\cdot$ & $4.22$& $5.87$ & $\cdot$   \\
\hline
\textbf{Prediction time}&& & $3.40$& $2.10$& $2.42$ & $3.05$ \\
\hline
\end{tabular}
\end{adjustbox}
\end{table}

\section{Temperature-related mortality in Italy}
%\subsection{Descriptive analysis}
As the simulation study showed that results are rather robust to the choice of partitioning units, we partitioned Italy according to province boundaries. Mainland Italy consists of $93$ provinces, different in the number of municipalities included, ranging from $6$ to $312$ with a median number of $60$. To avoid fitting submodels to very small partitions, provinces containing fewer than $25$ municipalities were merged with adjacent provinces. This resulted in $85$ core partitions, containing between $27$ and $312$ municipalities (Figure S2, Supplementary Materials). To diminish boundary effects, we extended each region with its second order neighbourhood. Running the $85$ partitions sequentially, computation time was approximately 12,5 hours.

Northern and Southern municipalities are highly different in terms of temperature distribution, as shown by Figure S3 in the Supplementary Materials. Moreover, the municipality-specific median number of reported deaths is zero for $96\%$ of the municipalities, but some serious outliers can be identified (Figure S4, Supplementary Materials).

\begin{figure}
    \centering
    \includegraphics[width=\linewidth]{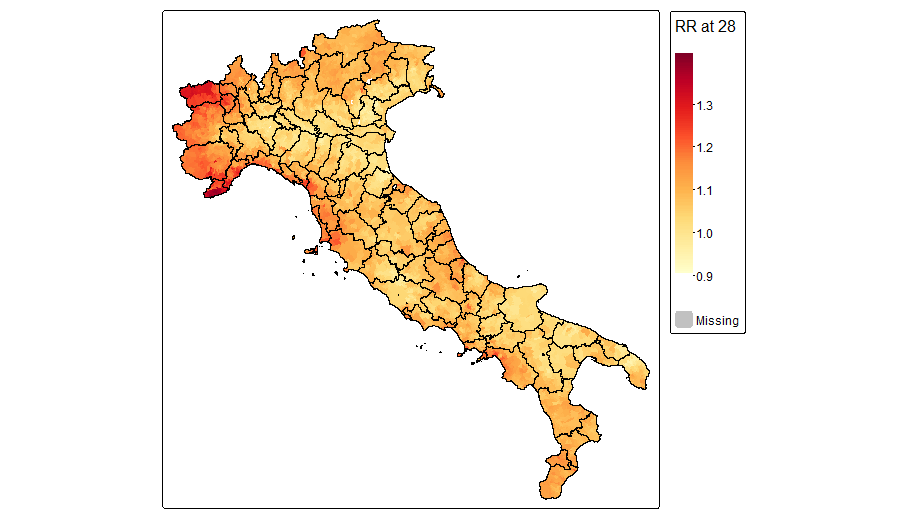}
    \caption{\emph{Estimated overall RR at a temperature of $28$ degrees Celsius, compared to a reference of $20$ degrees Celsius, for the $7122$ Italian municipalities, overlaid with province boundaries.}}
    \label{fig:italy_k2}
\end{figure}

\begin{figure}
    \centering
    \includegraphics[width=\linewidth]{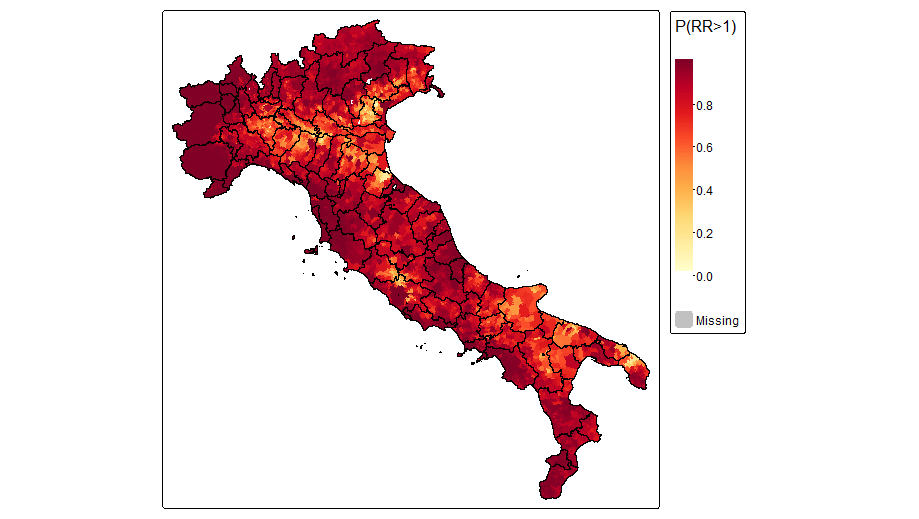}
    \caption{\emph{Exceedance probability that overall RR at $28$ degrees Celsius, compared to $20$ degrees Celsius, is larger than 1.}}
    \label{fig:exc_prob}
\end{figure}

%\subsection{Temperature-mortality relationship}
Figure~\ref{fig:italy_k2} presents the posterior mean of the overall cumulative RR at $28$ degrees , relative to $20$ degrees, summed over lags 0–8 days for every Italian municipality. The estimates indicate substantial spatial variation across Italy, with comparatively higher risk in the North-western regions and in several coastal municipalities. Because posterior means alone do not convey estimation uncertainty, Figure~\ref{fig:exc_prob} shows the posterior exceedance probability that the cumulative RR at $28$ degrees exceeds $1$. Elevated heat-related mortality risk was supported by high posterior probabilities in large parts of Italy, although the strength of this evidence varied geographically.

\begin{figure}
    \centering
    \includegraphics[width=\linewidth]{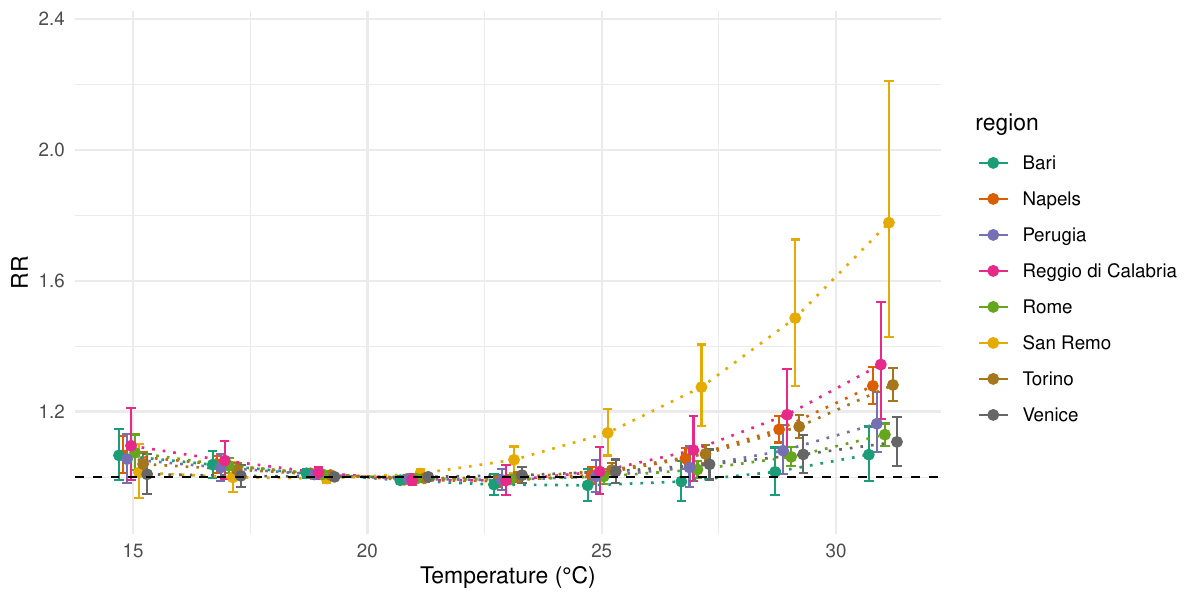}
    \caption{\emph{Estimated overall RR at different temperature values, together with the credible intervals, for $8$ selected Italian cities.}}
    \label{fig:RR_cities}
\end{figure}

To investigate the shape of the temperature–mortality association over a broader temperature range, we selected $8$ cities spread over the country, i.e. San Remo, Torino, Rome, Venice, Naples, Bari, Perugia and Reggio di Calabria (Figure S5, Supplementary Materials). Figure~\ref{fig:RR_cities} shows the estimated overall RR across temperature values, together with their credible intervals.  The estimated curves were similar at lower temperatures but diverged at higher temperatures. The RR curves for Bari, Perugia, Rome and Venice remain relatively flat, whereas the relative risk increased more strongly at high temperature values in several other municipalities. Rome, Bari and Perugia are municipalities with the highest average maximum temperature amongst the selected municipalities. Hence, this trend may indicate that the population of these cities is better adapted to extreme heat events, compared to other cities. Although Venice is not part of this list, it has a unique landscape, exposed to constant marine ventilation. The presence of open water and sea breezes may diminish the severe impact of heat on the body \citep{Peng2025}. 
In contrast, San Remo showed the steepest increase above approximately 25°C, although its estimates were also relatively uncertain. San Remo has the lowest average maximum temperature amongst the selected cities, which may indicate poor adaptation. Moreover, San Remo has an old population, with approximately $12\%$ of the inhabitants aged above $70$ in $2026$ \citep{citypopulation_sanremo}.

This analysis shows heterogeneity in the temperature-mortality relationship amongst the Italian cities. The divide-and-conquer approach made it possible to estimate municipality-specific temperature–mortality associations for all 7,122 municipalities, despite the computational infeasibility of fitting a single global model. To assess the sensitivity of the number of neighbours used to expand the regions, we repeated the analysis using first and third order neighbourhoods. The resulting relative-risk estimates were broadly consistent with those obtained using the second order neighbourhood specification, indicating limited sensitivity to this modelling choice (see Supplementary Materials, Figure S6-S7). Furthermore, comparison with an analysis using $75$ random grid partitions, containing between $34$ and $148$ municipalities, instead of province boundaries, revealed a similar trend as well (Figure S8).

\section{Conclusion}
Spatially-varying DLNMs are a flexible technique to allow for location-specific exposure-lag-response curves, but their computational requirements limit their application to datasets with only a few hundred areas. In big-data problems, memory and time constraints pose significant challenges, requiring a more scalable approach. In this paper, we adapted the `divide-and-conquer´ method to the spatially-varying DLNM framework, enabling the analysis of fine-scale nationwide spatio-temporal data that could not be fitted using a single global model. 

The simulation studies showed that the proposed method preserved an accuracy comparable to a global model in terms of RMSE. The method is relatively robust to both the choice of partitioning structure and the number of partitions. While increasing the number of partitions reduces the computational burden of each individual model, it also increases the total number of models to be fitted. Consequently, overall computation time reflects a trade-off between per-model complexity and the number of partitions. When different models can be fitted in parallel, computation time can be substantially improved. Its main advantage is therefore computational feasibility and scalability rather than shorter runtime.

The application of the `divide-and-conquer´ method to 7,122 municipalities in mainland Italy showed significant spatial heterogeneity in the temperature-mortality relationship. These differences may reflect heat adaptation, but also differences in demographic and socio-economic factors. This analysis therefore offers detailed insights that can generate hypotheses regarding underlying differences between municipalities. Such insights can be useful for refining public health guidelines and may support the design of more locally tailored heat-health interventions.

Several limitations of the proposed method should be acknowledged. First, partitioning the spatial domain might reduce the ability of the model to capture long-range spatial dependence, exceeding the partition boundaries, potentially leading to boundary effects. Although the use of a $k$-order neighbourhood mitigates this issue, residual discontinuities might remain. Second, although the simulation study and temperature-mortality study in Italy showed robustness against the choice of partition structure, sensitivity might remain to some extent. Third, fitting separate models across partitions can lead to increased uncertainty compared to a single model, particularly in regions with limited data. Fourth, although this method addresses the computational burden associated with big-data problems, the need to fit and combine different models introduces additional methodological and computational complexity.  The divide-and-conquer procedure introduces additional choices regarding partition construction, neighbourhood order, parallelization, and the combination of overlapping estimates.
Finally, the spatially-varying DLNMs are descriptive with respect to spatial effect modification. Possible explanations for the observed heterogeneity should therefore be investigated in future studies using municipality-level modifiers.

\section*{Supplementary materials}
\textbf{Online Appendix}: Contains additional information and figures. (Supplementary.pdf)\\

\section*{Acknowledgements}
The computational resources and services used in this work were provided by the VSC (Flemish Supercomputer Center), funded by the Research Foundation - Flanders (FWO) and the Flemish Government - department EWI.

\section*{Funding}
T.N. gratefully acknowledges funding from Research Foundation - Flanders (grant no. G0A3M24N).
\section*{Competing Interest Statement}
The authors have declared no competing interest.

\newpage
\newpage
\bibliographystyle{apalike}
\bibliography{References}

\end{document}